\documentclass[final,numberedheadings]{aipproc}
\layoutstyle{6x9}

\begin{document}

\newcommand{\commentpcf}[1]{}
\newcommand{\Vlsr}{V$_{\rm LSR}$}
\newcommand{\prism}{CLIC}
\newcommand{\Vhc}{V$_{\rm HC}$}
\newcommand{\deeg}{$^{\rm o}$}
\newcommand{\Bpar}{B$_{\rm  | |}$}
\newcommand{\cc}{cm$^{-3}$}
\newcommand{\kms}{km s$^{-1}$}
\newcommand{\HH}{H$_2$}
\newcommand{\nHI}{$n$(H$^{\rm o }$)}
\newcommand{\nHeI}{$n$(He$^{\rm o }$)}
\newcommand{\nH}{$n$(H)}
\newcommand{\nel}{$n$(e$^{\rm - }$)}
\newcommand{\NHI}{$N$(H$^{\rm o }$)}
\newcommand{\NHeI}{$N$(He$^{\rm o }$)}
\newcommand{\NDI}{$N$(D$^{\rm o }$)}
\newcommand{\NH}{$N$(H)}
\newcommand{\NOI}{$N$(O$^{\rm o }$)}
\newcommand{\NFeII}{$N$(Fe$^{\rm + }$)}
\newcommand{\FeD}{$N$(Fe$^+$)/$N$(D$^{\rm o}$)}
\newcommand{\FeII}{Fe$^{\rm + }$}
\newcommand{\AlII}{Al$^{\rm + }$}
\newcommand{\CII}{C$^{\rm + }$}
\newcommand{\CIIstar}{C$^{\rm + * }$}
\newcommand{\SiIII}{Si$^{\rm ++ }$}
\newcommand{\MgII}{Mg$^{\rm + }$}
\newcommand{\MgI}{Mg$^{\rm 0 }$}
\newcommand{\CaII}{Ca$^+$}
\newcommand{\NaI}{Na$^{\rm o }$}
\newcommand{\NNaI}{$N$(Na$^{\rm o }$)}
\newcommand{\NCaII}{$N$(Ca$^{\rm + }$)}
\newcommand{\OI}{O$^{\rm o }$}
\newcommand{\DI}{D$^{\rm o }$}
\newcommand{\R}{$N$(Fe$^{\rm + }$)/$N$(D$^{\rm o }$)}
\newcommand{\lya}{Ly-$\alpha$}
\newcommand{\cmtwo}{cm$^{-2}$}
\newcommand{\ntot}{$n_{\rm tot}$}
\newcommand{\chiH}{$\chi_{\rm (H)}$}
\newcommand{\HI}{H$^{\rm o}$}
\newcommand{\HeI}{He$^{\rm o}$}
\newcommand{\NeI}{Ne$^{\rm o}$}
\newcommand{\glong}{$l$}
\newcommand{\glat}{$b$}
\newcommand{\lbV}{$l$,$b$,$V$}

\newcommand\aj{{AJ}}% 
          % Astronomical Journal 
\newcommand\araa{{ARA\&A}}% 
          % Annual Review of Astron and Astrophys 
%\newcommand\apj{{ApJ}}% 
\newcommand\apj{{ApJ}}% 
          % Astrophysical Journal 
\newcommand\apjl{{ApJ}}% 
          % Astrophysical Journal, Letters 
\newcommand\apjs{{ApJS}}% 
          % Astrophysical Journal, Supplement 
\newcommand\ao{{Appl.~Opt.}}% 
          % Applied Optics 
\newcommand\apss{{Ap\&SS}}% 
          % Astrophysics and Space Science 
\newcommand\aap{{A\&A}}% 
          % Astronomy and Astrophysics 
\newcommand\aapr{{A\&A~Rev.}}% 
          % Astronomy and Astrophysics Reviews 
\newcommand\aaps{{A\&AS}}% 
          % Astronomy and Astrophysics, Supplement 
\newcommand\azh{{AZh}}% 
          % Astronomicheskii Zhurnal 
\newcommand\baas{{BAAS}}% 
          % Bulletin of the AAS 
\newcommand\jrasc{{JRASC}}% 
          % Journal of the RAS of Canada 
\newcommand\memras{{MmRAS}}% 
          % Memoirs of the RAS 
\newcommand\mnras{{MNRAS}}% 
          % Monthly Notices of the RAS 
\newcommand\pra{{Phys.~Rev.~A}}% 
          % Physical Review A: General Physics 
\newcommand\prb{{Phys.~Rev.~B}}% 
          % Physical Review B: Solid State 
\newcommand\prc{{Phys.~Rev.~C}}% 
          % Physical Review C 
\newcommand\prd{{Phys.~Rev.~D}}% 
          % Physical Review D 
\newcommand\pre{{Phys.~Rev.~E}}% 
          % Physical Review E 
\newcommand\prl{{Phys.~Rev.~Lett.}}% 
          % Physical Review Letters 
\newcommand\pasp{{Pub. Astron. Soc. Pacific}}% 
          % Publications of the ASP 
\newcommand\pasj{{PASJ}}% 
          % Publications of the ASJ 
\newcommand\qjras{{QJRAS}}% 
          % Quarterly Journal of the RAS 
\newcommand\skytel{{S\&T}}% 
          % Sky and Telescope 
\newcommand\solphys{{Sol.~Phys.}}% 
          % Solar Physics 
\newcommand\sovast{{Soviet~Ast.}}% 
          % Soviet Astronomy 
\newcommand\ssr{{Space~Sci.~Rev.}}% 
          % Space Science Reviews 
\newcommand\zap{{ZAp}}% 
          % Zeitschrift fuer Astrophysik 
\newcommand\nat{{Nature}}% 
          % Nature 
\newcommand\iaucirc{{IAU~Circ.}}% 
          % IAU Cirulars 
\newcommand\aplett{{Astrophys.~Lett.}}% 
          % Astrophysics Letters 
\newcommand\apspr{{Astrophys.~Space~Phys.~Res.}}% 
          % Astrophysics Space Physics Research 
\newcommand\bain{{Bull.~Astron.~Inst.~Netherlands}}% 
          % Bulletin Astronomical Institute of the Netherlands 
\newcommand\fcp{{Fund.~Cosmic~Phys.}}% 
          % Fundamental Cosmic Physics 
\newcommand\gca{{Geochim.~Cosmochim.~Acta}}% 
          % Geochimica Cosmochimica Acta 
\newcommand\grl{{Geophys.~Res.~Lett.}}% 
          % Geophysics Research Letters 
\newcommand\jcp{{J.~Chem.~Phys.}}% 
          % Journal of Chemical Physics 
\newcommand\jgr{{J.~Geophys.~Res.}}% 
          % Journal of Geophysics Research 
\newcommand\jqsrt{{J.~Quant.~Spec.~Radiat.~Transf.}}% 
          % Journal of Quantitiative Spectroscopy and Radiative Trasfer 
\newcommand\jatp{{J.~Atmosph.~Terres.~Phy.}}% 
          % Journal Atmospheric and Terrestrial Physics
\newcommand\memsai{{Mem.~Soc.~Astron.~Italiana}}% 
          % Mem. Societa Astronomica Italiana 
\newcommand\nphysa{{Nucl.~Phys.~A}}% 
          % Nuclear Physics A 
\newcommand\physrep{{Phys.~Rep.}}% 
          % Physics Reports 
\newcommand\physscr{{Phys.~Scr}}% 
          % Physica Scripta 
\newcommand\planss{{Planet.~Space~Sci.}}% 
          % Planetary Space Science 
\newcommand\procspie{{Proc.~SPIE}}% 
          % Proceedings of the SPIE 
\title{The Solar Galactic Environment }
\author{Priscilla C. Frisch}{address={University of Chicago, Department of Astronomy and Astrophysics,5640 South Ellis Avenue, Chicago, IL  60637, USA } }
%\author{<author2>}{
%  address={<common address for author2 and author3>}
%}
%%%%%%%%%%%%%%%%%%%%%%%%%%%%%%%%%%%%

\begin{abstract} 
Combined heliosphere-astronomical data and models
enrich our understanding both of effects the solar galactic environment 
might have on the inner heliosphere, and of 
interstellar clouds.  Present data suggest that \FeD\ increases
toward the upwind direction of the cluster of interstellar
clouds (\prism) flowing past the Sun.  Cloud kinematics and abundances 
suggest an origin related to a supershell around the Scorpius-Centaurus Association.
The solar space trajectory indicates the Sun entered the \prism\ gas
relatively recently.  \end{abstract}

\maketitle

%%%%%%%%%%%%%%%%%%%%%%%%%%%%%%%%%%%%%%%%%%%%
%% MAINMATTER
%%%%%%%%%%%%%%%%%%%%%%%%%%%%%%%%%%%%%%%%%%%%

\section{Introduction}\label{sec:intro}

Combined studies of the heliosphere and surrounding interstellar cloud
provide data on the interstellar medium (ISM) both at a single
location (scale $\sim$500 AU, the heliosphere entry regions) and
averaged over sightlines toward nearby stars (scale $\sim$10$^6$ AU).
\commentpcf{Radiative transfer effects normally hinder the
interpretation of low column density clouds (\NH$<$10$^{19}$) such as
surrounds the Sun, but are probed naturally in heliosphere-ISM
studies.}  Major quests of this relatively young discipline include
finding the chemical composition, physical properties, galactography,
\footnote{ Here I introduce a new word ``galactography'' (the Galactic analog of ``geography''), which
refers to the natural features (e.g. physical properties, morphology)
of a region of the Milky Way Galaxy.}  isotropy, and homogeneity of
the \prism.  The ultimate goal is to determine whether historic
variations in the boundary conditions of the heliosphere may have
affected the inner heliosphere regions and climate of Earth.  The
astrophysical basis of these questions is addressed in this review.

In our Galactic neighborhood ($<$400 pc) interstellar clouds with
densities $10^{\rm -5} - 10^{\rm 5}$ \cc, temperatures $10 - 10 ^6$ K,
and velocities 0--150 \kms\ are detected.  The types and properties of
typical clouds include giant molecular clouds ($P$=(\nHI, \nel,
temperature, velocity)=($\sim$5000 \cc, $\sim$2 \cc, $\sim$20 K,
$\sim$0 \kms)), CO/\HH\ clouds ($P$=($\sim$500, $\sim$0.1, $\sim$50,
$\sim$0)), cold neutral clouds ($P$=($\sim$15, $\sim$0.15, $\sim$100,
$\sim$2)), warm neutral clouds ($P$=($\sim$15,$\sim$0.20, $\sim$3000,
$\sim$10)), warm partially ionized clouds such as the LIC
($P$=($\sim$0.2,$\sim$0.1, $\sim$6000, $\sim$15--20)), HII regions
($P$=($\sim$0, $\sim$10, $\sim$10$^4$, $\sim$0)), intermediate velocity
clouds ($P$=($\sim$0, $\sim$0.3, $\sim$8000, $\sim$50)), high velocity
clouds ($P$=($\sim$0, $\sim$0.5, $\sim$8000, $>$100)), and tenuous
soft-Xray emitting plasma ($P$=($\sim$0, $\sim$0.0005, $\sim 10^6$, unknown)).
The intrinsic composition of the ISM is uncertain because of the mass
tied up in dust grains ($\sim$1\%), and abundance variations between
cold and warm clouds which persist after correction for ionization 
\cite{SavageSembach:1996,Welty23:1999,Cartledgeetal:2001,SlavinFrisch:2002,FrischSlavin:2003}.

The \prism\ appears to be typical warm diffuse low density
interstellar material.  
Warm low density clouds fill $\sim$50\% of the galactic disk plane,
and hence are the most likely types of clouds to be encountered by the
Sun.  Also, warm low density clouds are generally detected at higher
velocities than cold clouds, although this may be partially a line
blending effect.  Observations of the 21 cm line indicate that
$\sim$60\% of the \HI\ is warm (500--13000 K), and $\sim$50\% of the
warm material is thermally unstable \cite{HeilesTroland:2003II}.  This
``not strongly absorbing'' 21 cm emission is distinctly associated
with superbubble shells \cite{Heiles:1980}.  The cold gas itself is
found in thin sheets with a width-to-thickness aspect ratio of $\sim
300$ \cite{HeilesTroland:2003II}.  Hydrogen H$\alpha$ recombination
emission from diffuse warm ionized gas indicates $\sim$10--30\% of the
ionized gas is associated with low density, warm (\nHI=0.2--0.3 \cc,
$T \sim$8000 K) rims of low column density neutral clouds ($<$2 x
10$^{20}$ \cmtwo) \cite{ReynoldsTufteHeiles:1995}.  If the LIC is a
partially ionized cloud rim, the denser cloud may be the neutral
material seen upwind toward $\lambda$ Sco.  When low frequency
(0.1--10 MHz) radio data are combined with the H$\alpha$ and pulsar
dispersion measurements, a consistent picture emerges indicating the
warm ionized gas consists of clumped low density material, $T \sim$7000
K, $<$\nel$> \sim$0.23 \cc, and $\sim$2 pc radius
\cite{PetersonWebber:2002}.  The Sun is $required$ to be located in
such a clump to explain the low frequency turnover in the radio
spectrum.  UV data also sample the same material, and warm ($\sim$7000 K)
low density plasma (\nel$\sim$0.3 \cc) is seen at low velocity
toward $\mu$ Col and at high velocity toward 23 Ori
\cite{Howketal:1999,Welty23:1999}. 
These generic properties for warm low density gas, the global association
between 21 cm emitting warm \HI\ gas and superbubble shells, and the
\prism\ dynamics and abundances (\S \ref{sec:origin}), suggest a
superbubble origin for the local ISM.

\section{The Nearest ISM}\label{sec:near}

\begin{figure}[!b] 
\includegraphics [width=3.0in]{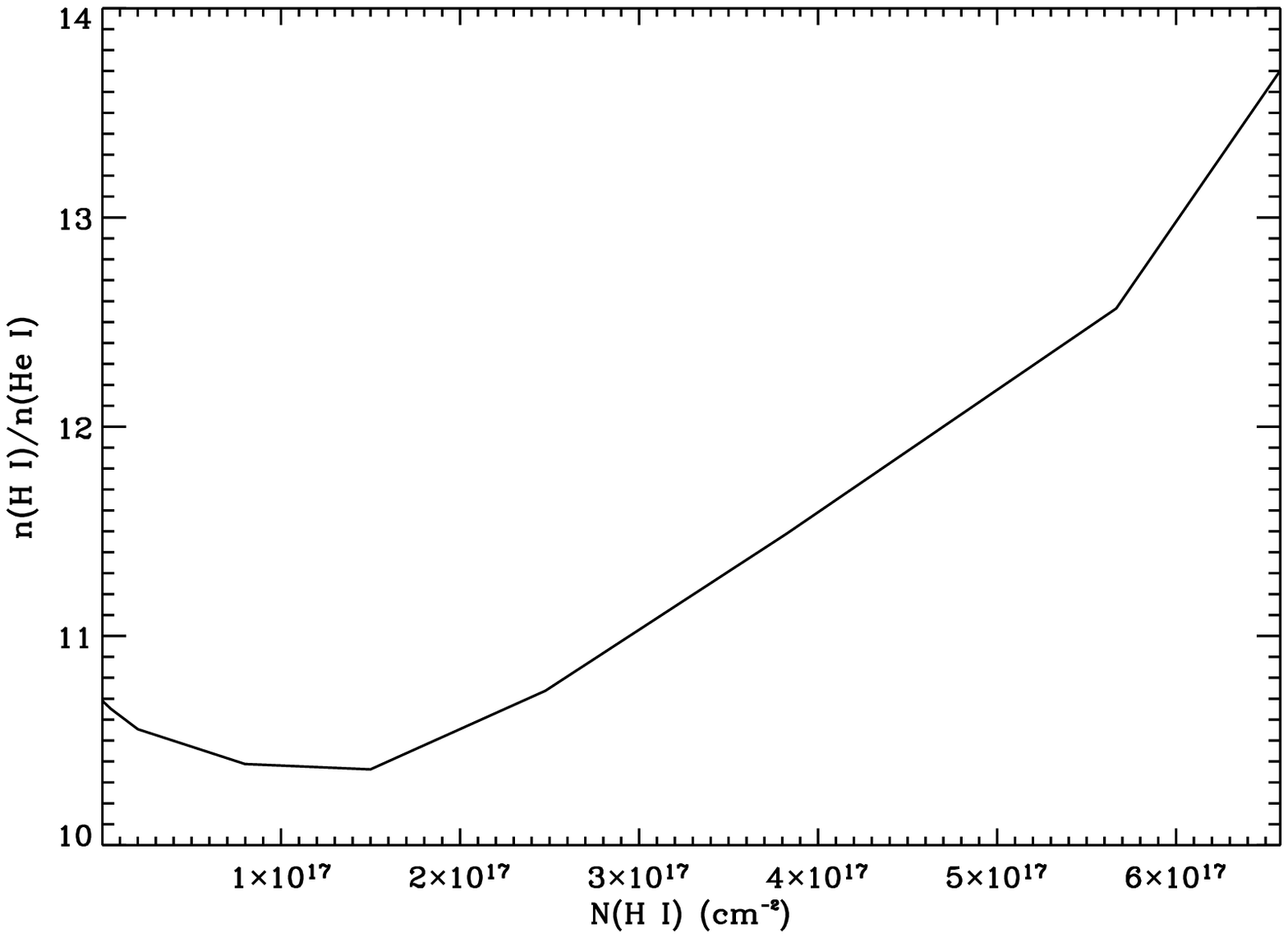}
\includegraphics [width=3.0in]{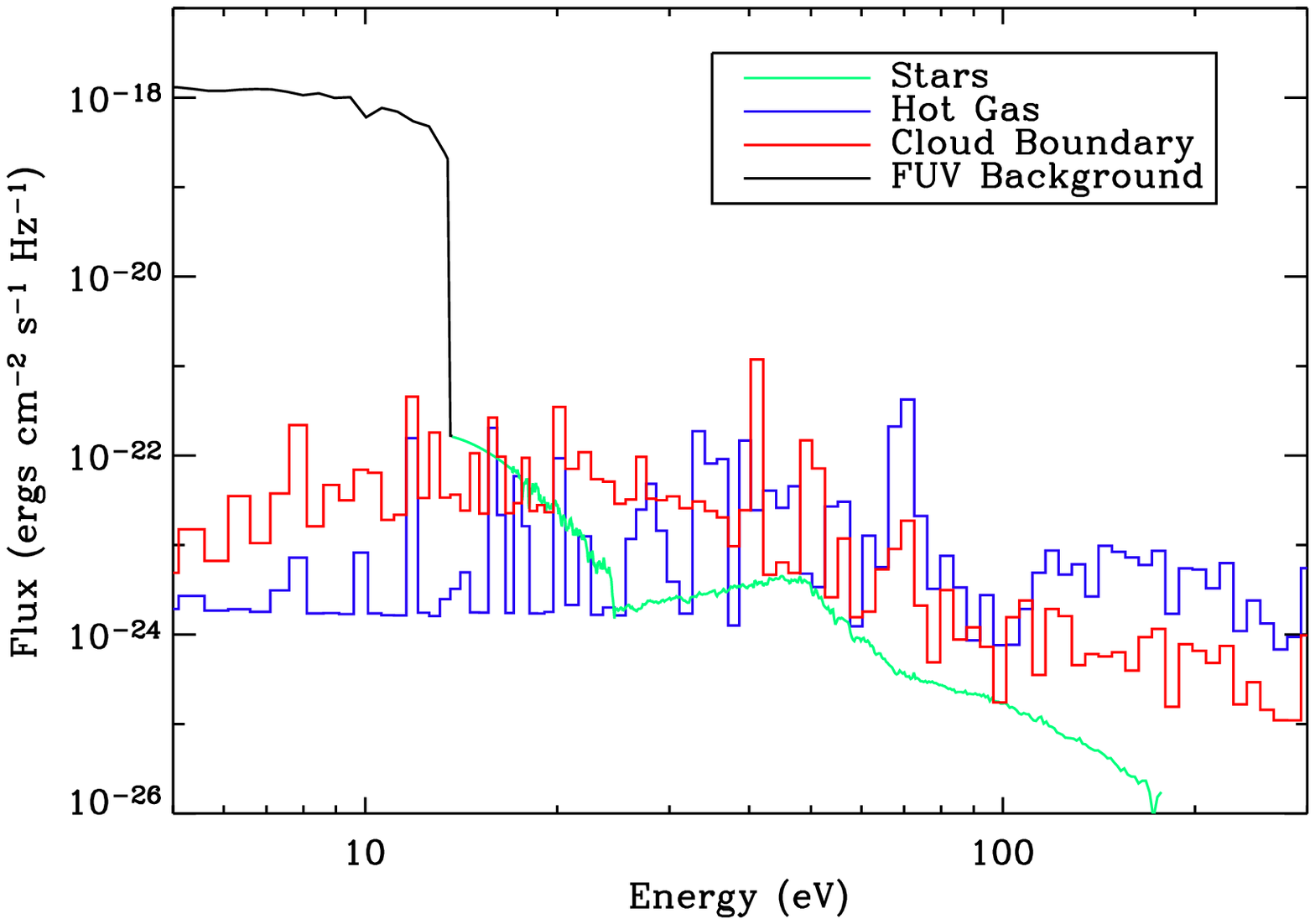}
\caption{{\bf Left:} The ratio between the volume densities of \nHI\ and
\nHeI\ as a function of distance from the LIC surface.  The Sun is
located at \NHI=6.5 x 10$^{17}$ \cmtwo.  {\bf Right:} The interstellar
radiation field at the cloud surface after dereddening.  Shown are
Model 2 results from \cite{SlavinFrisch:2002}.  }
\label{fig:HHe}
\end{figure}

Radiative transfer models of interstellar gas within $\sim$2 pc of the Sun, 
constrained by ISM inside the heliosphere and toward nearby stars, indicate that the LIC
at the solar location (SoL) has \nHeI$\sim$0.015 \cc, \nHI$\sim$0.22 \cc, \nel$\sim$0.1
\cc, and $\sim$70\% solar abundances
\cite{SlavinFrisch:2002,FrischSlavin:2003} (also see the paper by
Slavin in this volume).  Although these models yield LIC properties
that are now better constrained than for any other interstellar cloud,
many uncertainties remain.  Low column density clouds such as the LIC
are highly inhomogeneous because photons efficient for ionizing \HI\
($\sim$13.6 eV) are attenuated sharply in the outer cloud layers.  For
instance $n$(\HI)/$n$(\HeI) varies by $\sim$50\% in the LIC, since
\HI\ and \HeI\ are attenuated to 1/e at log \NHI$\sim$17.2 and 17.7
\cmtwo\ respectively (Fig. \ref{fig:HHe}).  Partially ionized elements
(H, He, Ar, Ne, O, N) and the ratios \MgII/\MgI\ and \CII/\CIIstar\
\footnote{The ratios \MgII/\MgI\ and \CII/\CIIstar\ are sensitive to
photoionization and collisional processes, respectively.}  provide the
most stringent constraints on \nel.  Observations of nearby stars give
sightline integrated results while ISM data inside of the solar system
(\HeI, \HI, pickup ions and anomalous cosmic rays) anchor the
radiative transfer models with data at the SoL.  Ionizations predicted
at the SoL are $\sim$ 29\% for H, $\sim$27\% for O, $\sim$48\% for He,
$\sim$41\% for N, $\sim$88\% for Ne, and $\sim$80\% for Ar.  The best
RT models reproduce \prism\ data toward $\epsilon$ CMa and yield
filtration factors (the fraction of interstellar neutrals successfully
penetrating the heliosheath regions) for O, N, Ar, He, and Ar that are
consistent with theoretical models \cite{Frisch:2003jgr}.

The RT models investigated by Slavin and Frisch (\cite{SlavinFrisch:2002,FrischSlavin:2003}, SF02) allow a range of H and He
ionization levels and yield \NHI/\NHeI=9--14 for LIC-like clouds;
this range is consistent with white dwarf data giving
\NHI/\NHeI=12--16 \cite{Vallerga:1996}.  The key parameter is the
radiation field, shown at cloud surface for Model 2 (SF02,
Fig. \ref{fig:HHe}).  The observed radiation field consists of
contributions from the diffuse soft X-ray background, massive stars
such as $\epsilon$ CMa, and white dwarf stars.  The radiation field must be
dereddened to infer the radiation field throughout the LIC from 
observations at the SoL.  Radiation with $\lambda \sim$ 500 A which
ionizes \HeI\ and \NeI\ is the least well constrained observationally.
A conductive interface on the cloud surface will contribute emission
in this interval (see the Slavin paper).  The principal deficiency of
the best RT models (chosen to match the widest range of observational data)
is the predicted temperature (which is $\sim$2000 K too high).  However since the
temperature is extremely sensitive to the abundance of the primary
coolant \CII, which has abundance uncertainties of $\sim$100\%
\cite{GryJenkins:2001}, this shortcoming may not be significant.

%%Figure 1

The interstellar magnetic field at the SoL is poorly understood, and
the picture is essentially unchanged since my 1990 talk at the Warsaw
COSPAR Symposium \cite{Frisch:1990}.  Data on the polarization of
nearby stars show a patch of magnetically aligned dust grains in the
upstream direction \cite{Tinbergen:1982}, but these results have not
been reproduced \cite{Leroy:1993A}.  From these data the magnetic
field direction is parallel to the galactic plane and directed toward
\glong$\sim$70\deeg.  This orientation is consistent with Voyager 3 kHz
radio emission data which show a preferred direction parallel to the
galactic plane \cite{KurthGurnett:2003}.  Possibly the 3kHz emission
is related to the draping of the magnetic field around the heliosheath
since the classical charged interstellar dust grains polarizing the
starlight (radii $\sim$0.1--0.2 $\mu$m) are deflected around the
heliosheath \cite{Frischetal:1999,LindeGombosi:2000}.  
Polarization strengths vary little in the first 10 parsecs, indicating the
polarization occurs within 10 pc of the Sun \cite{Frisch:1990}.  The
region of maximum starlight polarization is slightly below the
ecliptic plane ($\beta = 0 ^{\rm o} \rightarrow -15^{\rm o}$) but also
near the upstream direction in the local standard of rest (LSR)
\cite{Frisch:2003jgr}.  Weak field strengths ($\sim$1.5--3.0 $\mu$G)
are consistent with both polarization data \cite{Frisch:1990} and
Faraday rotation caused by ambient low density weakly ionized ISM
\cite{RandKulkarni:1989}.  Flux freezing arguments do not appear
valid, since Zeeman splitting studies of the 21 cm \HI\ line show that the
parallel component of the magnetic field has little
correlation with cloud density \cite{HeilesTroland:2003}.

\section{The \prism} \label{sec:origin} 

The kinematics and abundance patterns of the \prism\ tell us a lot
about its origin, in the first case because cloud motions are traced,
and in the second because gas-phase abundances are enriched by grain
processing in interstellar shocks \cite{Jones:1994,Jones:1996}.
Diffuse clouds are identified from the velocity structure of the
absorption lines, assuming Doppler broadening from thermal motions
plus a catch-all factor interpreted as turbulence.  Good data are
important for identifying velocity structure since component crowding
in velocity space increases exponentially with improvements in
instrumental resolution \cite{Welty23:1999}.  Observations of UV and
optical lines indicate \prism\ temperatures are mainly in the range
5000 K to $<$30000 K, with typical turbulent broadening of
$\xi$=1.5--2.5 \kms.  However several local colder clouds (T$<< 10^3 $
K) detected toward stars at distances of $\ge$50 pc may be part of the
Loop I shell \cite{FGW:2002}.

The Sun is embedded in a flow of interstellar cloudlets (the
\prism); it is a robust result that the upstream direction of this
flow is toward the Scorpius-Centaurus Association (SCA)
\cite{FGW:2002}.  Observations\footnote{Only the radial velocity component of the velocity is observed.}  of $\sim$100 absorption lines sampling
the \prism\ give a flow velocity with respect to the Sun
(heliocentric) of (\lbV) = (12.4\deeg, 11.6\deeg, --28.1$\pm$4.6 \kms).
This velocity vector can be converted to the LSR by removing the solar apex motion,
giving (\lbV)= (331.4\deeg,--4.9\deeg,--19.4 \kms) and (\lbV)=
(2.3\deeg,--5.2\deeg,--17.0 \kms), respectively, for the Standard and
cool star apex motions \cite{FGW:2002}.  \footnote{The Standard solar
apex motion is 19.7 \kms\ toward \glong=57\deeg, \glat=+22\deeg\
\cite{Allen:1973}, while the apex motion determined from Hipparcos
observations of cool stars is 13.4 \kms\ toward \glong=27.7\deeg,
\glat=32.4\deeg\ \cite{DehnenBinney:1998}.  The cool-star apex motion
reflects the greater age of cool stars which have kinematically
relaxed with respect to shorter-lived massive stars.}  The large
dispersion in the \prism\ velocities ($\sim$5 \kms) indicates a
turbulent flow.  Regardless of the assumed solar apex motion, the
upstream direction is toward the SCA \cite{Frisch:1981,FGW:2002},
which shows highly structured shell-like \HI\ filaments and a
superbubble (Loop I) with radius $>$90\deeg\ that dominates the
northern sky \cite{deGeus:1992}.

The LIC is a cloudlet in the \prism\ and it is the only cloudlet with
an unambiguous 3D velocity vector because of the Ulysses \HeI\ data
\cite{Witte:2004}.  The LIC velocity\footnote{The LIC heliocentric upstream vector is
(\lbV)= (3.3,+15.9,--26.3 \kms), which converts to an LSR vector
(\lbV)= (317.8 \deeg,--0.5\deeg,--20.7  \kms) for the Standard, and to
(\lbV)= (346.0\deeg,+0.1 \deeg, --15.7 \kms) for the cool star apex motion.}
projects to --17.3 \kms\ in the
$\alpha$ Cen (1.3 pc) sightline (using Ulysses \nHeI\ data
\cite{Witte:2004}), compared to observed \FeII, \MgII, \DI, and \AlII\
velocities ranging between --17.6$\pm$1.5 \kms\ and --19.6$\pm$0.6
\kms\ \cite{LinskyWood:1996,Woodetal:2001,RedfieldLinsky:2004}.  This
small velocity difference has been interpreted to mean the LIC
terminates within $\sim$10$^4$ AU of the Sun in this direction, but
the difference is smaller than the turbulent velocity of 1.5$\pm$0.3
\kms. 

At least five individual clouds are kinematically distinct in this
flow, including a cloud at --32 to --35 \kms\ (depending on the
angular extent of the cloud) and located within 5 pc of the Sun in the
solar apex direction
\cite{Lallementetal:1986,FGW:2002,Frisch:2003apex}.  A velocity
component denoted the "G-cloud" (for galactic center hemisphere) has been
identified to explain a large number of velocity components in the
upstream direction, including the cloud toward $\alpha$ Cen ($\sim$1.3
pc) \cite{Lallement:1996}.  The G-cloud, however, appears to result
from the overlap in velocity space between distinct cloudlets
clustering around the bulk flow velocity of \prism\ and at different
distances \cite{FGW:2002}.  However if the G-cloud is a single
cloudlet, the data shows it must be dense (\nHI$>$5 \cc,
\cite{Frisch:2003apex}).

\begin{figure}[!b]
 \resizebox{17pc}{!}{\includegraphics[angle=-90]{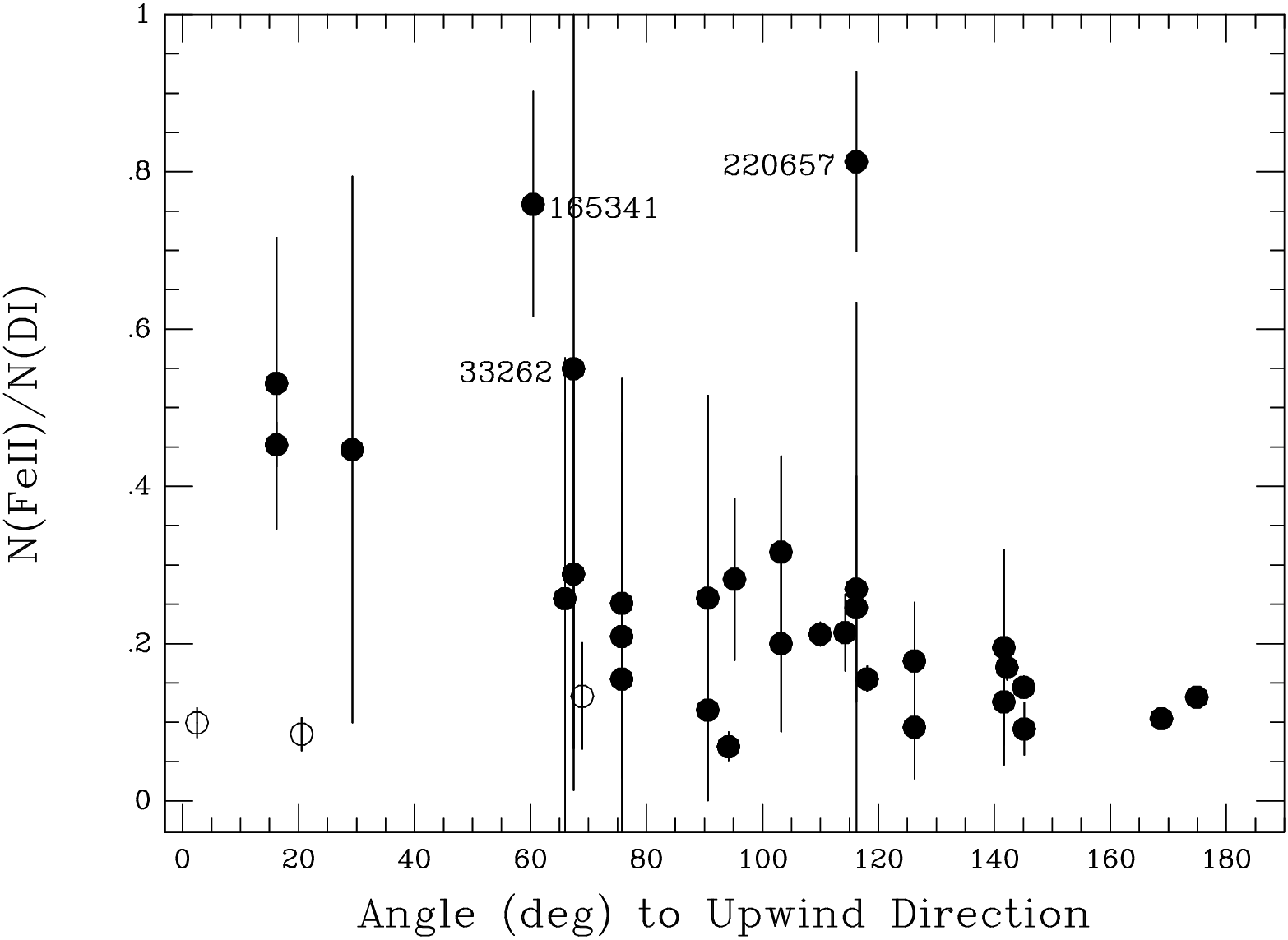}}
 \resizebox{17pc}{!}{\includegraphics[angle=-90]{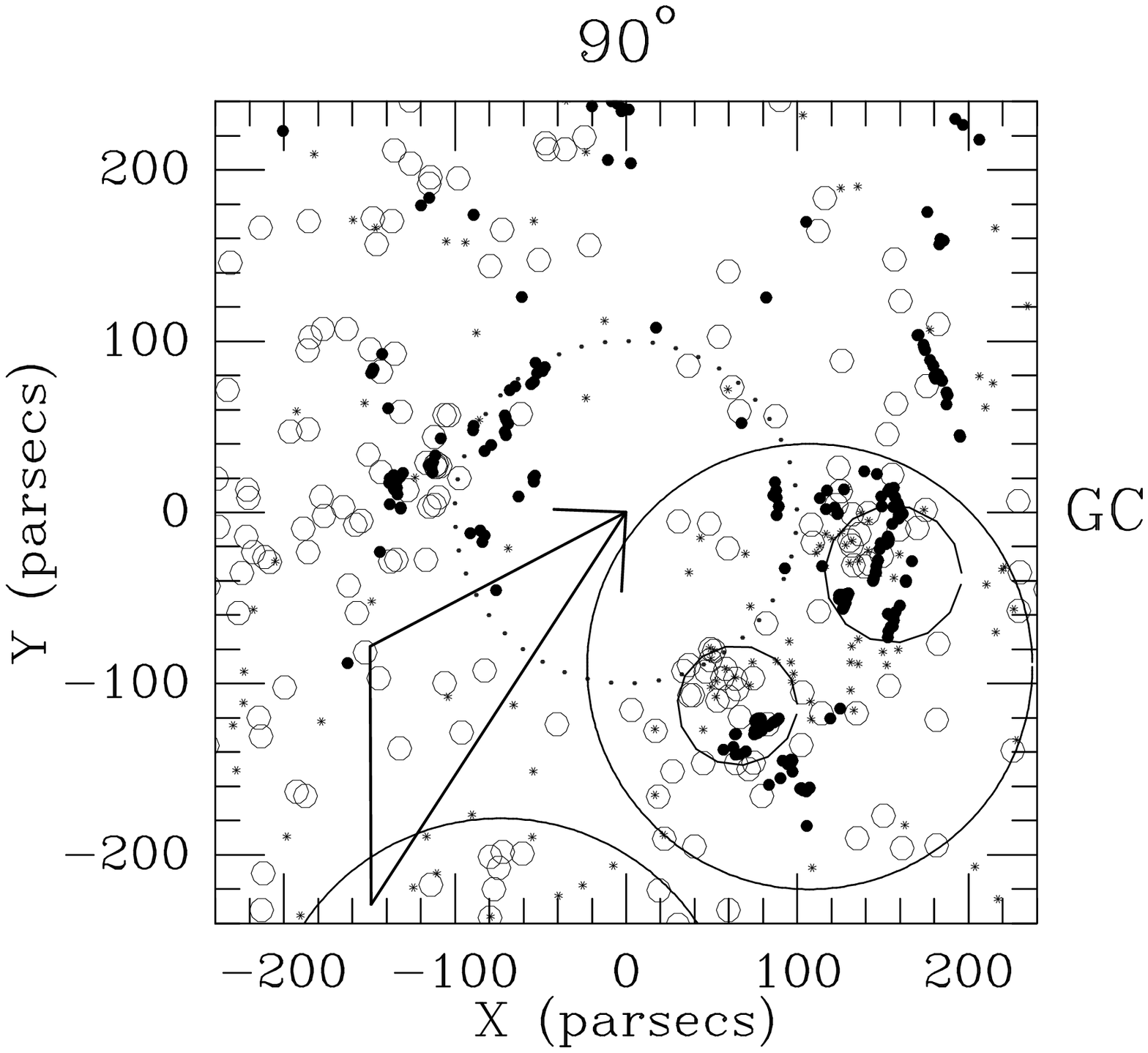}}
\caption{
{\bf Left:} 
\NFeII/\NDI\ plotted against the angle (in degrees) between the background 
star and the LSR upwind direction (using the Standard solar apex motion).  
The plot is based on data from
\cite{RedfieldLinsky:2002,RedfieldLinsky:2004,York:1983,FrischYorkFowler:1987,Lehneretal:2003}.
The G191-B2B component at --9.9 \kms\ appears at an angle 174\deeg, and
\NFeII/\NDI=2.14, and is not plotted.  The open circles represent
lower resolution data ($\sim$15--20 \kms) toward the stars $\alpha$
Oph, $\lambda$ Sco and HD 149499B.  The remaining data were acquired
primarily at resolutions 2--3 \kms.  For several stars, \NDI\ is
derived from \HI\ using \NDI/\NHI=1.5 x 10$^{\rm -5}$ or O/H=400 ppm.
{\bf Right:} 
The space trajectory of the Sun with respect to nearby
interstellar clouds within $\sim$200 parsecs.  Filled dots represent
CO and dust clouds from the compilation in \cite{DutraBica:2002}, open
circles represent infrared-bright (60 $\mu$m) dust surrounding B stars
\cite{GaustadVanBuren:1993}, and small dots are nearby early type
stars (O--B2).  Solar motion is shown for both Standard and cool-star
solar apex motions (arrows, see \S \ref{sec:origin}), and a dotted
circle is drawn at 100 pc.  The solid circles are the superbubbles
around the Upper Scorpius, Lower Centaurus-Crux and Upper Centaurus
Lupus subgroups of the SCA \cite{deGeus:1992,Frisch:1995}.  The
past deficit of ISM surrounding the Sun is evident. 
} \label{fig:FeD}
\end{figure}

Inhomogeneities in the ISM were first observed over 50 years ago as
systematic increases of \NCaII/\NNaI\ with cloud velocity
\cite{RoutlySpitzer:1952}.  Similar increases in refractory element
(e.g. Fe, Mg, Si, Mn) abundances with cloud velocity are
successfully modeled by the shattering and evaporation of interstellar
dust grains in shocks
\cite{Jones:1994,Jones:1996,SavageSembach:1996,Welty23:1999}.
Abundances of \FeII\ and \MgII\ in the \prism\ are reproduced by
thermal sputtering of dust grains in $\sim$100 \kms\ shocks
\cite{Frischetal:1999}.  However \FeII\ abundances are nonuniform
locally, and there is a factor of $\sim$6 difference between the
maximum value of the ratio \FeII/\DI\ in the upstream direction
(toward $\alpha$ Cen A,B and 36 Oph) and minimum value toward the
downstream direction (toward \glong$\sim$160\deeg\ and $\alpha$ Aur,
e.g. \cite{LinskyWood:1996,RedfieldLinsky:2004}).  Fig. \ref{fig:FeD}
(left) shows \FeD\ for absorption components in $\sim$30 stars
sampling the \prism\ plotted against the angle between the star and
LSR upstream direction (\glong=331\deeg, \glat=--5\deeg).  The \FeII\
abundance increases as the view-direction sweeps upwind, which is
consistent with either an ionization or abundance gradient in the
\prism.  The alternative possibility is that the Sun is near the
boundary of two cloudlets, as Collier et al.  \cite{Collieretal:2004}
have suggested based on data showing two interstellar neutral gas
streams at 1 AU.  Some components that are exceptions to this trend
may sample non-\prism\ gas.  
Components toward HD 333262 and HD 220657 appear to have either
anomalously large Fe abundances or ionization.  The nearby cloudlets
toward $\lambda$ Sco and $\alpha$ Oph are neutral and significantly
denser (\nHI$>$2--5 \cc) than the LIC
\cite{York:1983,Frisch:2003apex,Muelleretal:2004}.

%%Figure 2

\paragraph{Origin and Galactography} The local origin of the weak
\CaII\ absorption components observed toward Scorpius and Ophiuchus
stars has long been recognized
\cite{MunchUnsold:1962,Herbig:1968,MarschallHobbs:1972},
and identified with the interstellar material inside of the solar
system \cite{AdamsFrisch:1977}.  Based on cloud kinematics and
refractory abundances, Frisch
\cite{Frisch:1981,Frisch:1995,Frisch:1996} modeled the local gas as a
fragment of superbubble shell around the Scorpius-Centaurus
association that has expanded away from SCA region into the low
pressure interarm region of the Local Bubble.  Combined superbubble
shell expansion models and stellar evolution in the SCA subgroups
indicates the \prism\ properties are consistent with a 4 Myr old
superbubble shell\footnote{In a typographical error,
\cite{Frisch:1995} incorrectly lists this age as 400000 years.}
expanding into low density gas remaining from an earlier episode of
star formation.  Building on the superbubble concept, the \prism\ has
been modeled as a Rayleigh-Taylor instability formed by the
interaction of Loop I and the Local Bubble, which were speculated to
have merged 1-10 Myrs ago \cite{Breitschwerdtetal:2000}.  These instabilities involve magnetic
reconnection in bubble walls which eject $\sim$1 pc diameter blobs of
gas to the solar vicinity.  A third
model of the \prism\ origin origin assumes that magnetic flux tubes
detach from the walls of the Local Bubble around Sun and spring back
toward Sun carrying clumps of gas embedded in relatively strong
magnetic fields ($\sim$7 $\mu$G, \cite{CoxHelenius:2003}), but
predict local magnetic field directions conflicting with the polarization
data.

%%Figure 3
\begin{figure}[!th]
 \resizebox{18pc}{!}{\includegraphics{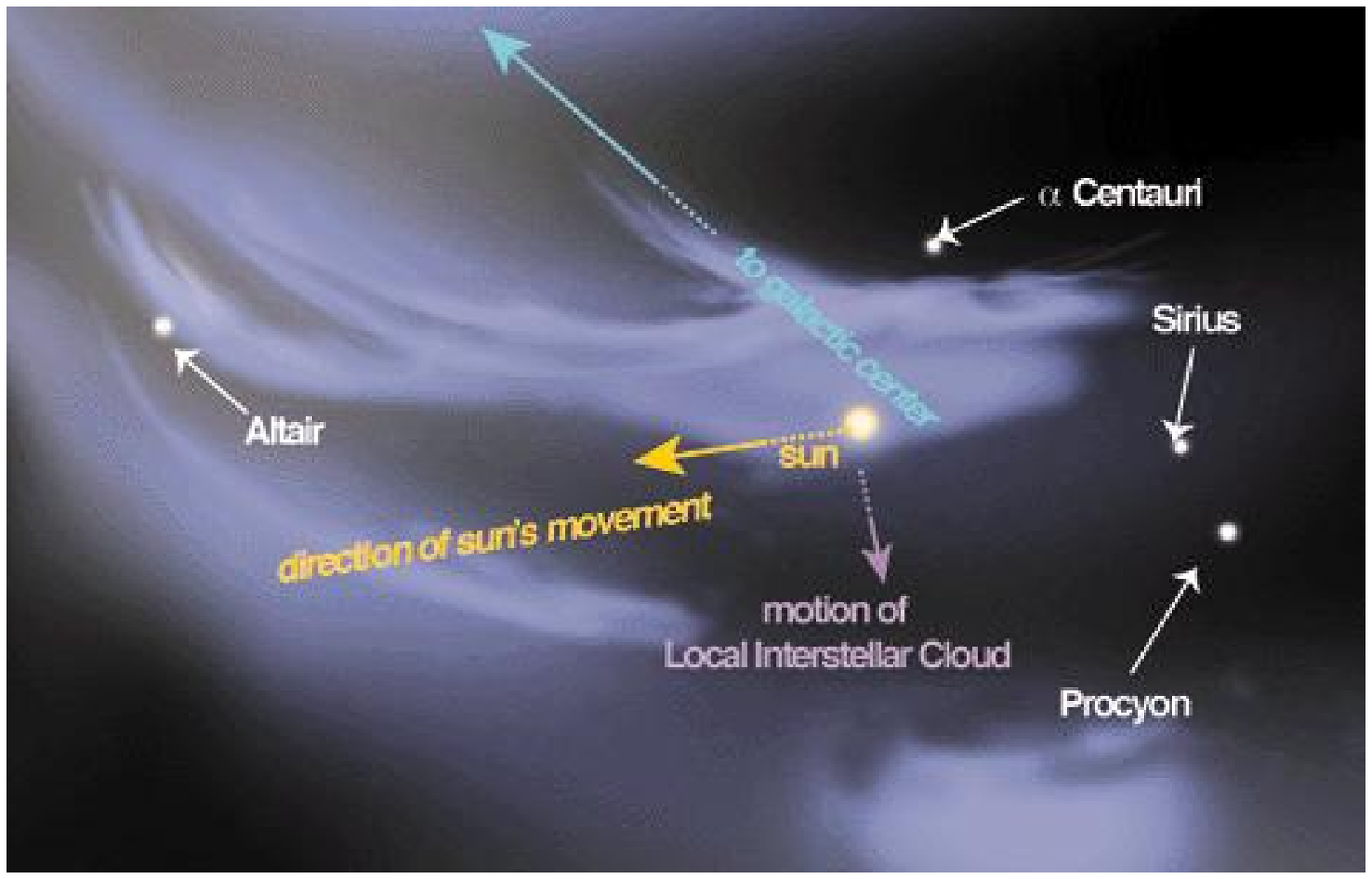}}
\caption{The local ($<$5 pc) ISM filamentary structure shown
here is based on the assumption that the LIC velocity vector is 
parallel to the cloud surface normal \cite{Frisch:1994,Frisch:1996}, which 
then yields a magnetic field direction nearly parallel to the 
cloud surface (based on the polarization data of \cite{Tinbergen:1982}).
Figure credit:  Linda Huff at American Scientist and \cite{Frisch:2000amsci};
used with permission.}
\label{fig:amsci}
\end{figure}

Tracing the solar trajectory backwards in time indicates the Sun has
been embedded in the low density interior of the Local Bubble for
several million years \cite{FrischYork:1986,Frisch:1997}.  The solar trajectory
compared to dust clouds in the Local Bubble is shown in Fig.
\ref{fig:FeD} (right), for both the Standard and cool-star apex
motions.  

I prefer local galactography models which invoke filamentary or
sheet-like structures for the \prism, rather than egg-like models
based on smoothing algorithms \cite{Linskyetal:2000}.  The ISM is
replete with spectacular filamentary structures, found at all
velocities and in both neutral and ionized gas.  Emission lines trace
relatively young supernova blast waves interacting with low density
clouds, such as features seen toward the Cygnus Loop where
collisionless shocks accelerate the plasma but not neutral gas
\cite{PatnaudeFesen:2002}.  
Starlight polarization \cite{Tinbergen:1982}, the EUV source
distribution \cite{Warwicketal:1993}, and the high column densities
inferred for the G-cloud toward $\alpha$ Oph, $\lambda$ Sco and HD
149499B \cite{Frisch:2003apex,York:1983,Lehneretal:2003} all suggest
that the \prism\ material in the upstream direction is within $\sim$10
pc of the Sun.  

For the morphology shown in Fig. \ref{fig:amsci}, the Sun emerged from 
the Local Bubble interior and entered the outflow from the SCA within
the past $\sim 10^5$ years, and the LIC within the past 2000--8000
years \cite{Frisch:1994,Frisch:1996}.  Edge effects occurring as the Sun passes
through/between cloudlets where magnetic field strength or density may vary
appear to perturb the heliosphere and modify the cosmic ray flux
in the inner heliosphere, and possibly the terrestrial climate
\cite{Frisch:1997,Frischetal:2002b,Florinskietal:2003}.
The emergence of {\it homo sapiens} coincided approximately in time
with the Sun's entry into the CLIC, which resulted in a contracted heliosphere
and the appearance of anomalous cosmic rays
\cite{Frisch:1999,Muelleretal:2001}.  Cosmic rays at the Earth appear to affect
cloud formation \cite{Carslawetal:2002}
and the global electrical circuit \cite{Roble:1991}.

\begin{theacknowledgments}
The author would like to thank NASA for research support through the
grants NAG5-11005 and NAG5-13107 to the University of Chicago.
\end{theacknowledgments}

%\bibliographystyle{aipprocl} % if natbib is missing
%\bibliography{master,frisch}

%%%%%%%%%%%%%%%%%%%%%%%%%%%%%%%%%%%%%%%%%%%
%% Just a reminder that you may have to run bibtex
%% All of it up to \end{document} can be removed
%% if you don't like the warning.
%%%%%%%%%%%%%%%%%%%%%%%%%%%%%%%%%%%%%%%%%%%
\IfFileExists{\jobname.bbl}{}
 {\typeout{}
  \typeout{******************************************}
  \typeout{** Please run "bibtex \jobname" to obtain}
  \typeout{** the bibliography and then re-run LaTeX}
  \typeout{** twice to fix the references!}
  \typeout{******************************************}
  \typeout{}
 }

\end{document}